\documentclass[twocolumn]{aastex6}

\usepackage{ amssymb }
 \usepackage{multirow}
\usepackage{times}
\usepackage{ amssymb }
\usepackage{color}

\slugcomment{Draft version}

\begin{document}

\title{An active, asynchronous companion to a redback millisecond pulsar}

\author{Andr\'{e} D. van Staden}
\affil{11 Sonopstreet, Bredasdorp, Western Cape, 7280, South Africa}

\and 

\author{John Antoniadis\altaffilmark{1}}\footnote{Corresponding author; \hyperref[antoniadis@dunlap.utoronto.ca]{antoniadis@dunlap.utoronto.ca}} 
\affil{Dunlap Institute for Astronomy \& Astrophysics, University of Toronto, 50 St. George Street, Ontario, M6P2R7, CA}



\begin{abstract}

PSR\,J1723$-$2837 is a ``redback'' millisecond pulsar (MSP) with a low-mass  companion in a 14.8\,h orbit. The system's properties closely resemble those of ``transitional'' MSPs that alternate between spin-down and accretion-powered states. 
In this paper we report on long-term photometry of the 15.5\,mag companion to the pulsar. 
We use our data to illustrate that the star experiences sporadic activity, which we attribute to starspots. 
We also find that the companion is not tidally locked and infer  $P_{\rm s}/P_{\rm b}= 0.9974(7)$ for the ratio between the rotational and orbital periods. 
Finally, we place constraints on various parameters, including the irradiation efficiency and pulsar mass. 
We discuss similarities with other redback MSPs and conclude that  starspots may provide the most likely explanation for the often seen irregular and asymmetric optical lightcurves.  
\end{abstract}


\section{Introduction} \label{sec:intro}
Targeted radio searches towards Fermi-LAT unclassified point sources have unravelled numerous  millisecond pulsars (MSPs) with non-degenerate  companions \citep[][]{rob13}. Their defining properties include radio eclipses around superior conjunction (i.e. when the pulsar is behind the companion), $\gamma-$ and X$-$ray emission, rapid orbital-period modulations \citep{akh+13}, and strong optical variability.  
This rich phenomenology suggests that the pulsar companions are losing mass, thereby providing a unique opportunity to probe accretion physics and stellar evolution in the presence of a strong external heating source. 

Eclipsing MSPs are commonly  classified  as either ``black widows'' or ``redbacks'', depending on 
the mass of the companion ($m_{\rm c}$ of up to few $10^{-2}$\,M$_{\odot}$, and  $m_{\rm c}\simeq 0.2-0.7$\,M$_{\odot}$ 
respectively). 
 Even though the origin and evolution of these systems remain ambiguous \citep{cct+13,bdh+14}, 
the relative population sizes suggest that the two subclasses are not directly linked, but rather represent stable endpoints of distinct evolutionary paths \citep{cct+13}. This  argument 
is somewhat  challenged by recent observations illustrating that redback MSPs can 
transition between a spin-down radio MSP state and a low-mass X$-$ray binary (LMXB) state \citep{asr+09}. 

State changes have now been observed in three binaries, with recurrence time-scales of order  few years \citep{asr+09,pfb+13, sah+14,pah+14,bph+14,bpa+14, tya+14,rrb+15}. 
According to the prevailing theory, transitions are likely driven by irradiation feedback on the companion, which alters its size and mass-loss rate,  
triggering propeller-type instabilities on an accretion disk \citep[see discussion in][]{pah+14}.  

Indeed, for most known redback MSPs the companion luminosity  modulates  
strongly with the orbital period, presumably due to the varying view of a heated area on the stellar surface \citep[e.g.][]{bvr+13,dcm+14,bkb+16,rmg+15,drc+16}. 
Thus far however, interpretation of redback lightcurves has been proven challenging. For instance, models in which the companion's side facing the pulsar is directly heated by the pulsar wind do not adequately describe the  asymmetries and phase-dependence of the observed ``day/night'' variations \citep[e.g.][]{rmg+15,drc+16}. 

Among  alternatives, the most viable option appears to be a modified geometry in which  the pulsar wind is reprocessed by  intra-binary shocks, the presence of which is suggested by  phase-resolved {X}-ray observations. 
\cite{rs16} recently demonstrated that  one can introduce an arbitrary degree of asymmetries in the lightcurve, if the reflected radiation is modelled self-consistently. 
Their model seems to improve model fits substantially, at least in the case of PSR\,J2215+5135, a redback MSP with a lightcurve well-sampled in binary phase. 
Still, numerous questions remain. For instance, the intra-binary shock model does not provide a  explanation for the often seen secular changes in average flux. Also, its long-term self-consistency remains to be tested as, thus far, most redback lightcurves have been sampled over a  limited number of orbital cycles. 

Herein we present long-term optical  photometry of PSR\,J1723$-$2837, a 1.86\,ms redback MSP in a 14.8\,h orbit around a main-sequence-like companion. 
PSR\,J1723$-$2837 was discovered in the Parkes Multibeam survey in 2004 \citep{fsk+04}. In a long-term radio-timing study, \cite{cls+13} find that the system shares common properties with transitional MSPs, including irregular radio eclipses around superior conjunction and timing noise suggestive of tidal interactions between the pulsar and its Roche-lobe filling companion. The system has a bright ($V_{\rm mean}\simeq15.5$\,mag) optical counterpart, making it an ideal case-study for redback MSPs. 
We use our data to illustrate that the companion star experiences periods of increased surface activity, leading to variability modulated at the rotational period, which is not synchronized with the orbital period. This behaviour can fully account for the (transient) asymmetries seen over individual orbital cycles. 

\section{Observations} \label{sec:obs}
 PSR\,J1723$-$2837 was observed by the lead author between 2014 August 3 and  2015 October 27 from Overberg, South Africa. The location of the observatory allows for continuous monitoring of the target for up to 9 hours per night, or 60\% of the orbital period. The site suffers from minimal light pollution and the seeing typically ranges from 2 to 3 arcsec. 
 
All observations were conducted with a 30\,cm Cassegrain telescope 
and a commercial SBIG ST9e CCD camera cooled to $-15^{\rm o}$\,C, with 20\,$\mu$m$^2$ pixels arranged in a 512$\times$512 grid. 1844 images 
were collected over sixty seven nights, with exposure times ranging from 300\,s (in 2014) to 600\,s (in 2015).   No photometric filters were used.  Guiding corrections were performed with the integrated tracking CCD at 1\,sec intervals, always using the same star.

The images were reduced using standard dark and sky-flat frames acquired sporadically throughout the campaign. The  magnitudes reported below are based on differential ensemble-type aperture  photometry, computed with C-Munipac/Muniwin (v.\,2.0.17\footnote{\url{http://c-munipack.sourceforge.net/}}), a public photometry program. Fluxes were measured inside an aperture enclosing 2 times the FWHM of the local PSF, and the sky was extracted from a surrounding region up to 5 times the FWHM, but excluding the area close to our source, which contains a faint star. We experimented with different settings, each time extracting consistent results. 

A set of six isolated bright objects were selected to derive a virtual comparison star. We verified this choice by monitoring the stability for one of them against the other five. As can be seen in Figure\,\ref{fig:1}, the residual flux is consistent with white noise, with a root-mean-square value of 9\,mmag. 

Time stamps were derived from PC time which was synchronized to an international time server and  checked regularly against a GPS master clock. All measurements below refer to barycentric  times corresponding to the center of each exposure. 

Binary phases were calculated using the pulsar timing ephemeris of \cite{cls+13}, 
adopting $T_{\rm asc} =$MJD\,55425.320466, for the time of ascending node and $P_{\rm b} = 0.615436473$\,days, for the orbital period. We also account for the small secular change in the orbital period, $\dot{P}_{\rm b} = -3.50 \times 10^{-9}$, which could result in a delay of up to   $\Delta P_{\rm b} \simeq -0.5$\,sec over the span of the observations. Higher-order variations in $P_{\rm b}$ were omitted, as their effect is too small to influence our results \citep[see][]{cls+13}. 
Under the adopted convention, $\phi=0.25$ coincides with  the superior conjunction, when the pulsar is eclipsed by its companion. 

\section{Results}\label{sec:res} 
\begin{figure*}[t!]
\begin{center}
\includegraphics[width=\textwidth]{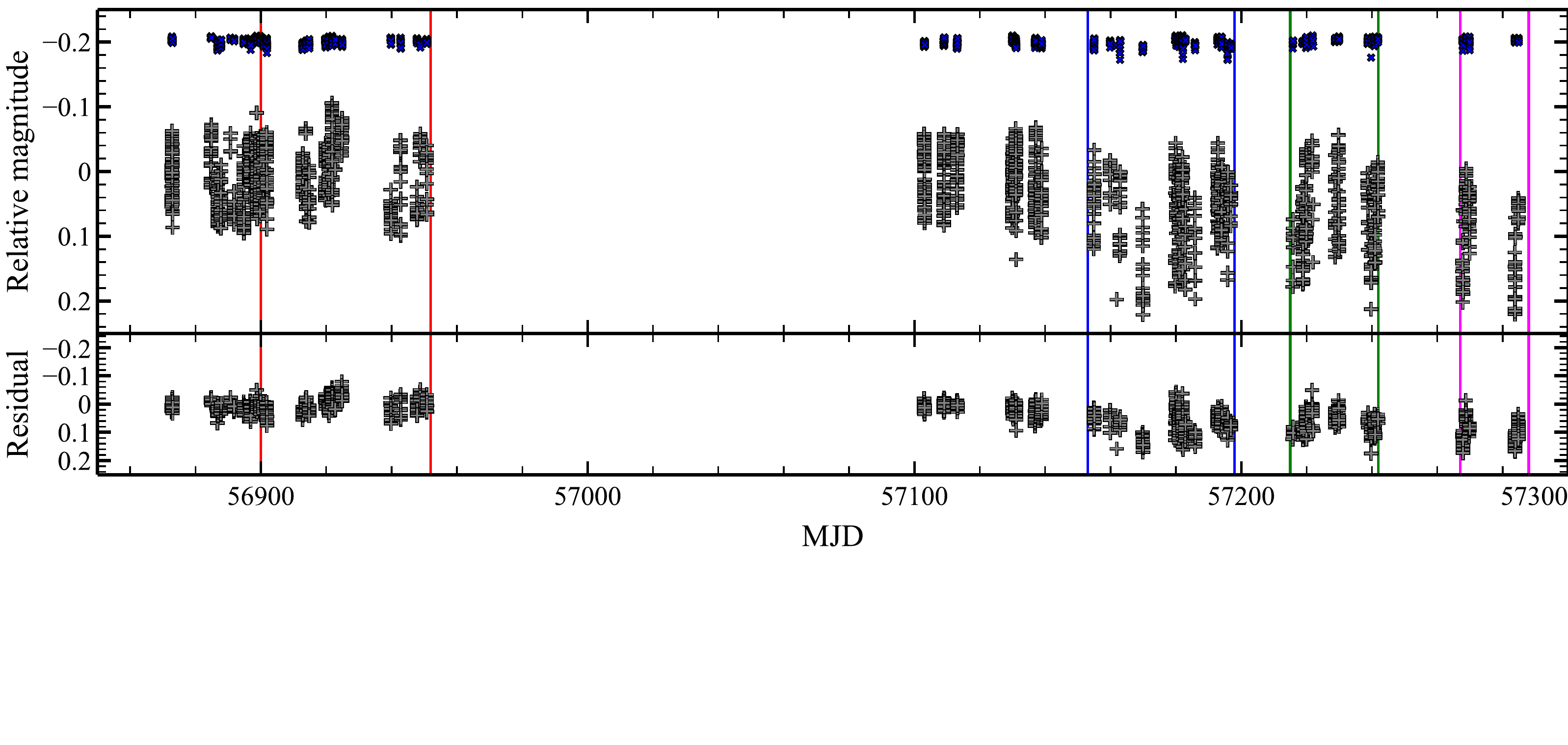}
\caption{Time-series photometry of PSR\,J1723$-$2837 from August 2014 to October 2015. The magnitude of the pulsar companion relative to that of a reference star is shown in black (shown in blue). The upper panel shows the raw lightcurve, while the lower panel shows the residual signal after subtracting the ellipsoidal modulation. The epochs marked with red, blue, green and magenta lines indicate the periods of increased surface activity discussed in the text. The folded lightcurves for these epochs are shown in Figure\,\ref{fig:2}.}
\label{fig:1}
\end{center}
\end{figure*}

\begin{figure*}[t!]
\begin{center}
\includegraphics[width=\textwidth]{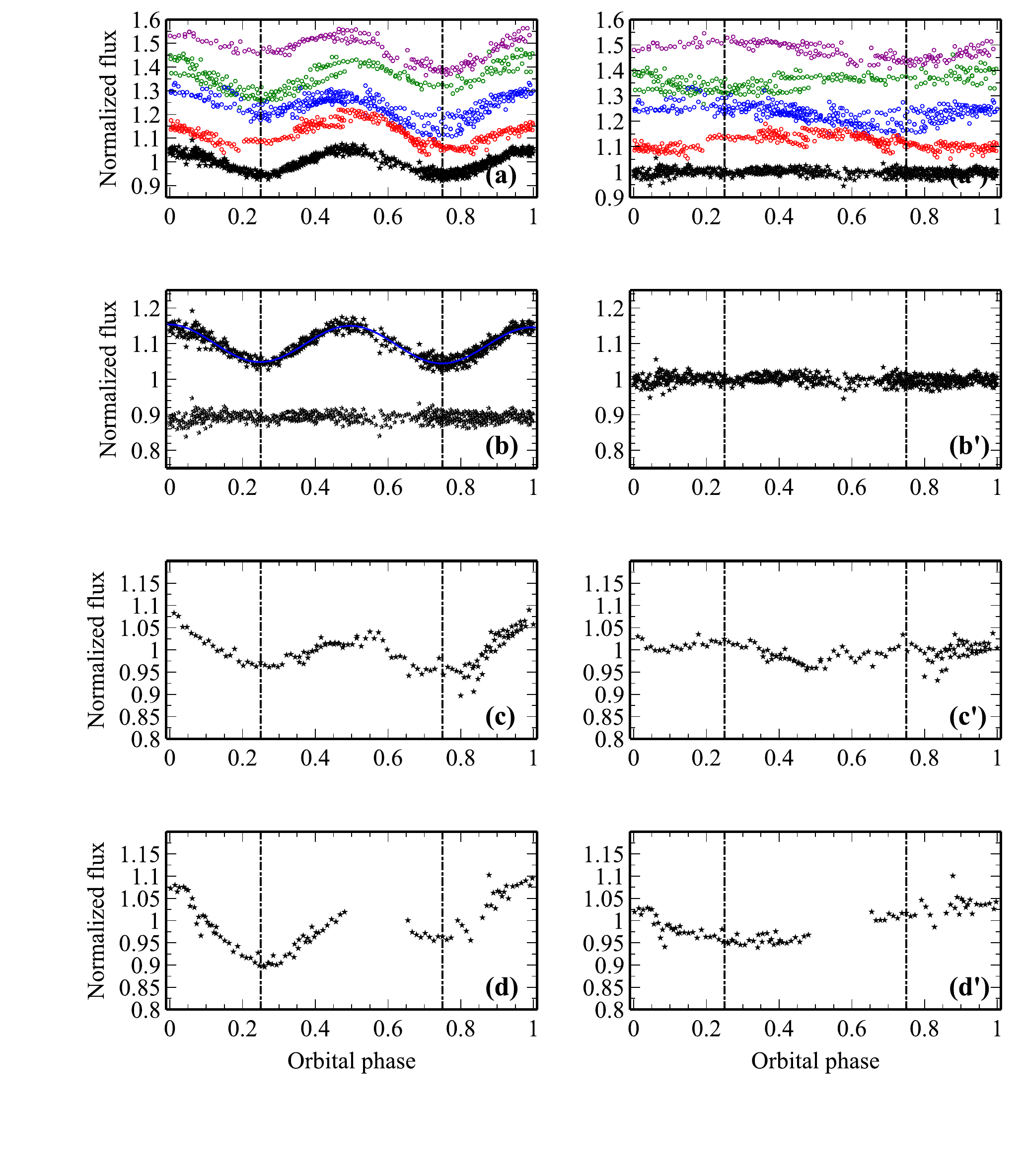}
\caption{The lightcurve of PSR\,J1723$-$2837 folded using the orbital ephemeris of \cite{cls+13}. The right column shows the residuals after subtracting the ellipsoidal signal inferred from the Lomb-Scargle periodogram. Panel a shows lightcurve sequences  for the quiescent period (black), and MJDs 56901$-$56950 (red), 57154$-$57197 (blue), 57215$-$57241 (green) and 57267$-$57288 (purple). 
Panels b shows the lightcurve during quiescence (as above), the best-fit synthetic lightcurve (in blue; see \S\ref{sec:dis}), and the residuals of the fit. Panels c \& d show instances of the lightcurve between MJDs\,57192--57197 and  MJDs\,57215--57221 respectively. The dashed vertical lines indicate the orbital phases corresponding to superior/inferior conjunction.}
\label{fig:2}
\end{center}
\end{figure*}

Figure\,\ref{fig:1} shows the lightcurve of the companion to PSR\,J1723$-$2837 from August\,2014 to October 2015. The most dominant trend is a $\Delta m\simeq 0.07$\,mag drop in flux after $\sim$MJD\,57160. 
In Figure\,\ref{fig:2}a we show our entire dataset folded with the orbital ephemeris of \cite{cls+13}. Here, one can already see a seemingly ``chaotic'' behaviour which would be difficult to attribute to a stable  geometric configuration.  We find that this is caused by periods of increased activity, intervened by periods of ``quiescence''. In what follows we discuss these separately and use them to constrain various system parameters. 
\subsection{Tidal deformation and direct irradiation}\label{sec:ellipsoidal}
Upon visual inspection, we identify two periods of quiescence, between MJDs\,56872--56901  and MJDs\,57102--57130. The folded lightcurve corresponding to these epochs (Figure\,\ref{fig:2}b) is dominated by a persistent signal modulated at the first  orbital harmonic. 
We model the lightcurve in two ways. First, we fit an empirical model which includes two sinusoids at the fundamental and first orbital harmonic (with phases fixed using the timing ephemeris), and a third degree polynomial for normalization. 

We infer fractional amplitudes of $\mathcal{A}_{\rm 1}=0.0031(9)$ and $\mathcal{A}_{\rm 2}=0.0507(9)$
 for the two sine waves, and polynomial coefficients consistent with zero, except for the constant normalization factor. 
Varying the polynomial degree (from 0 to 5) yields consistent results. When allowing for phase offsets we find $\Delta \phi_{\mathcal{A}_{\rm 1}} \simeq 0.3$ and a somewhat larger fractional amplitude $\mathcal{A}_{\rm 1}=0.0045(7)$  for the fundamental. 
The reduced $\chi^2$ for these fits is $\simeq 0.2$, most likely due to overestimated uncertainties.  Thus far, we were unable to derive more reliable measurement errors from our data. The rms scatter of the residuals suggests that true uncertainties are $\sim$2.5 smaller. However, in what follows we do not rescale the uncertainty estimates, to stay on the conservative side. 

We next turn to the Lomb-Scargle periodogram  
of the data \citep{lomb,scargle}, shown in Figure\,\ref{fig:3}. From  the aforementioned windows we infer  $\mathcal{A}_{\rm 2}=0.0527(12)$  for the strongest signal, peaking at $P_{\mathcal{A}_{\rm 2}}=7.38526(9)$\,h\footnote{All values and associated uncertainties derived from frequency analysis are based on Monte-Carlo simulations, where we generate several 10\,000 realizations of the lightcurve. We report mean values and their standard deviations. }, which is fully consistent with the prediction of the timing ephemeris. 
If we instead use the entire dataset we find; $\mathcal{A}_{\rm 2}=0.0532(8)$ (Figure\,\ref{fig:3}a). 
This  suggests that the true uncertainties are somewhat larger, and therefore we adopt,

\begin{equation}
\mathcal{A}_{\rm 2}=0.0512(25)
\end{equation}
for the rest of our analysis. 
At two times $P_{\mathcal{A}_{\rm 2}}$ we infer $\mathcal{A}_{\rm 1}\simeq0.003$ for the quiescent data, and $\mathcal{A}_{\rm 1}\simeq0.008$ for the full dataset, although, as we discuss in \S\ref{sec:spots} this is likely contaminated by a stronger peak in the vicinity.

We believe that the dominant source of variability is ellipsoidal variations from the tidally-distorted companion. If one treats these perturbatively, then $\mathcal{A}_2$ can be thought of as the linear term of the Fourier expansion describing the ellipsoidal signal, with $\mathcal{A}_2=f_{\rm ell}(R_{\rm c}/a)^3 q\sin^2{i}$ \citep{bee89,mn93}. Here $a$ is the semi-major axis of the companion, $q\equiv m_{p}/m_{c}$ is the mass ratio and $f_{\rm ell}$ (of order unity) depends on the properties of the atmosphere and the CCD response (see \S\ref{sec:params} for details). In \S\ref{sec:params} we use our measurement to constrain various parameters of the system. 

For $\mathcal{A}_{\rm 1}$ the interpretation is not as straight-forward. For direct 
irradiation one would expect $\mathcal{A}_1 \simeq f_{\rm irr}(T^{4}_{\rm irr}/32T^{4}_{\rm eff})\sin i$, where $T_{\rm irr}=L_{\rm psr}/4\pi \alpha^2 \sigma_{\rm B}$ is the 
irradiation temperature, $T_{\rm eff}$ is the effective photospheric temperature, and $f_{\rm irr}$ accounts  for the atmospheric albedo and the irradiation efficiency. In principle, the  dependence on inclination could be stronger, depending on the wind radial profile. 
If one takes these estimates  at face value then $f_{\rm irr}\simeq$4\%, adopting the system parameters discussed in \cite{cls+13} and below. 
However, we do not believe that our data support direct irradiation as firstly ,there is evidence for misalignment between inferior conjunction and maximum flux and secondly, the signal may be contaminated by starspots and aliases of the ellipsoidal modulation.

\subsection{Starspots and stellar rotation}\label{sec:spots}
We now focus on data collected during periods of increased activity. 
\begin{figure}[h]
\begin{center}
\includegraphics[width=0.45\textwidth]{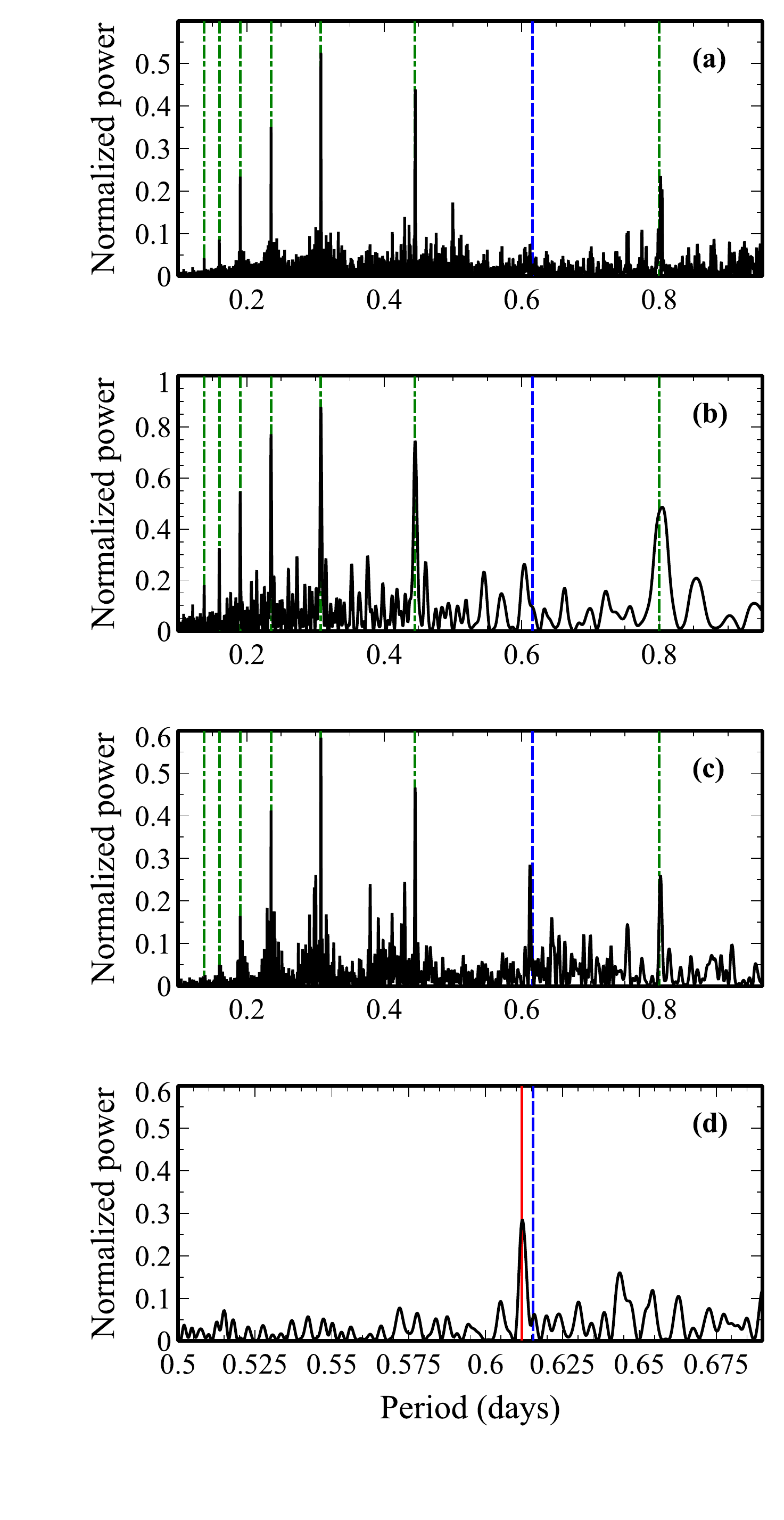}
\caption{Lomb-Scargle periodogram of our dataset. Dashed-dotted lines show the aliases of the ellipsoidal signal, while the dashed blue line is the orbital period. Panel\,a shows the spectrum of the entire dataset, Panels b and c are for the quiescent and active periods respectively (see text). Finally, panel d is the same as c, but focuses around the orbital period. The best-fit spin-period of the companion is shown in red.}
\label{fig:3}
\end{center}
\end{figure}
Figure\,\ref{fig:3}c shows the Lomb-Scargle periodogram of the data acquired between MJDs\,56901--56950, and after MJD\,57161.
Here, the most striking feature is an extra signal with fractional amplitude $\mathcal{A}_{\rm s} \simeq0.031$ at $P_{\rm s}=0.6118(4)$\,d. This is similar, but not identical, to the orbital period inferred from the timing solution, or self-consistently from our data using the ellipsoidal variations.  

\begin{figure}[h]
\begin{center}
\includegraphics[width=0.45\textwidth]{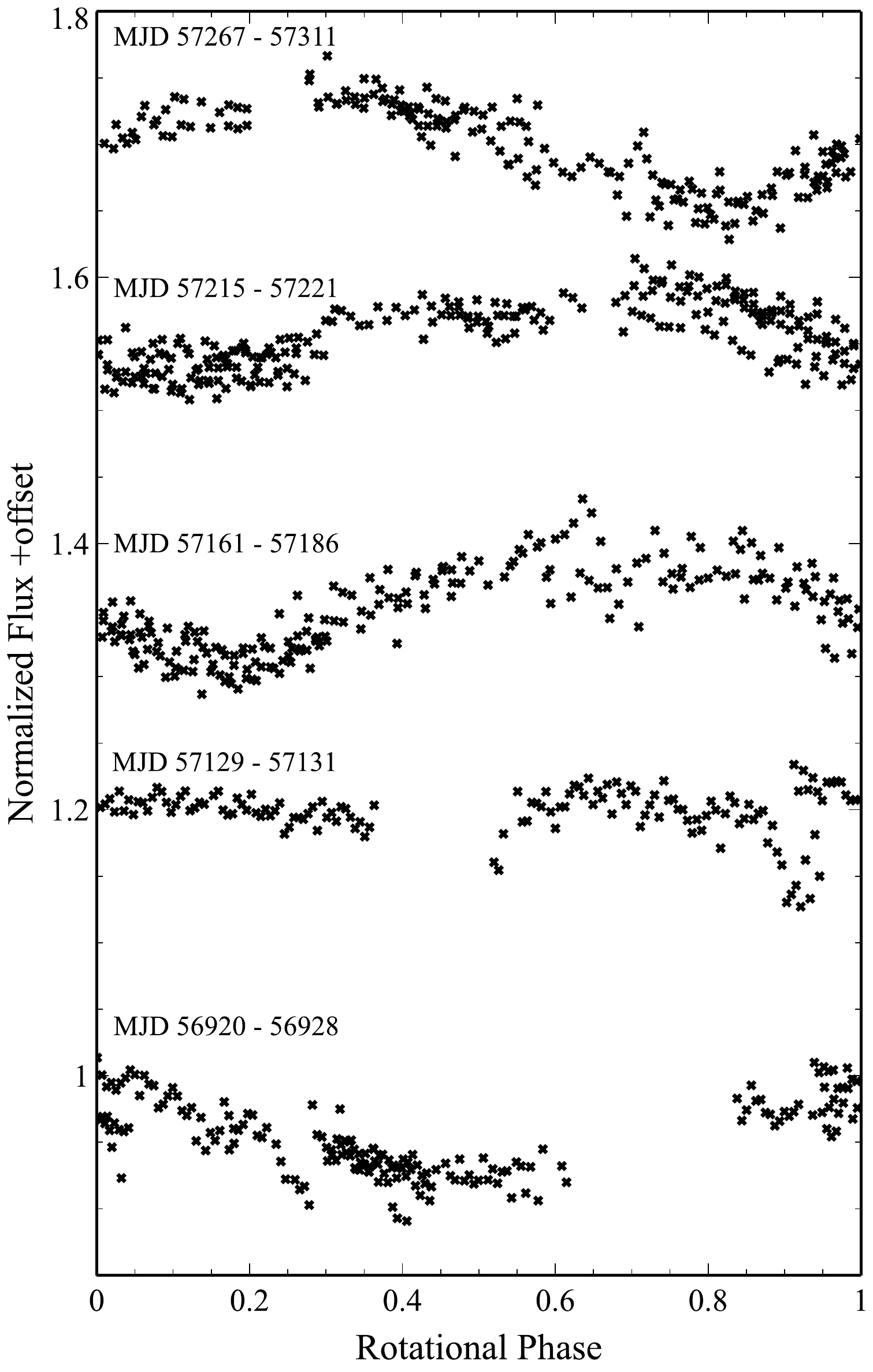}
\caption{Examples of lightcurve residuals folded at the rotational period inferred in\S\ref{sec:spots}. From bottom to top, each sequence is offset by 0.2 flux units from the previous.}
\label{fig:4}
\end{center}
\end{figure}

In Figure\,\ref{fig:4} we plot sequences of the residual signal folded at $P_{\rm s}$, after subtracting the ellipsoidal variation inferred in \S\ref{sec:ellipsoidal}. Interestingly, the 
shape of the residual evolves over time and occasionally deviates significantly from a sine. 
We believe that the most obvious explanation for this phenomenology is stellar activity, in which case $P_{\rm s}$ traces stellar rotation. 
Each sequence in Figure\,\ref{fig:4} must correspond to a distinct group of star-spots. Overall, we find evidence for $\sim$8 spots with lifetimes ranging between $\sim$3 and 45\,days. Generally, after the drop in average flux on $\sim$MJD\,57161, the  minima appear to be deeper and more persistent.

 Evidently, the signal does not maintain coherence over time and therefore our estimate for $P_{\rm s}$ above must be biassed. 
Repeating the frequency analysis for the three largest and most persistent  star-spots (the three upper sequences in Figure\,\ref{fig:4}) yields:
$P_{\rm s}^{1,2,3} = 0.61423(52),0.6148(52),0.6131(7)$\,days respectively, and a weight-averaged  spin-to-orbital-period ratio of, 
\begin{equation}
{P_{\rm s}}/{P_{\rm b}}=0.9974(7)
\end{equation}
using our self-consistent estimate for $P_{\rm b}$ from \S\ref{sec:ellipsoidal}. In averaging, we assume rigid rotation, given that the photometric precision and cadence do not suffice to  to resolve differential rotation. 
We further discuss this result in the following section. 
\section{Ramifications} \label{sec:dis}
To our knowledge, the data presented here establish the most densely-sampled photometric lightcurve  for any eclipsing binary pulsar to date,  also providing the first direct inference of the spin rate for any MPS  companion. 
As shown in \S\ref{sec:spots}, there is clear evidence for asynchronous rotation. Here we explore some of the ramifications of our measurements in more detail.

\subsection{System parameters}\label{sec:params}
PSR\,J1723$-$2837 hosts an MSP, and hence it is reasonable to expect that the system went through at least one phase of Roche-lobe overflow during which the companion was tidally locked.  If one assumes that the spin-up of the companion was caused by a recent  contraction, e.g. due to a reconfiguration of the internal magnetic field on a dynamical timescale \citep[][]{mw83,app92},  then the rotational rate inferred above relates directly to the Roche-lobe filling factor,
\begin{equation}
f_{\rm RL}\equiv R_{\rm c}/ R_{\rm RLO} \simeq (P_{\rm s}/P_{\rm b})^{1/2}= 0.9987(3).
\end{equation}
In practice, the tidal torque is acting to spin-down the star, while an additional magnetic torque may also influence the spin evolution. 
In any case, the torque-free assumption should yield a reliable upper limit for the filling factor. 

Adopting  $f_{\rm RLO}\simeq1$, one can use the constraints on ellipsoidal variations and $q$ to derive an estimate on the inclination and component masses, since $(R_{\rm c}/a)=f_{\rm RL}(R_{\rm RL}/a)$, where the latter is also a function of $q$ \citep{egg83}. Using $q=3.3(5)$ from \cite{cls+13} we infer $\sqrt{f_{\rm ell}}\sin i = 0.44(2)$; with the more precise estimate, $q=3.45(2)$ from our follow-up work (Antoniadis {\it et al.}, in prep.),   
\begin{equation}
\sqrt{f_{\rm ell}} \sin i = 0.438(12) 
\end{equation}
and, through the binary mass function \citep{cls+13}, $f_{\rm ell}^{3/2}m_{\rm c} =0.181(16)$\,M$_{\odot}$ and  $f_{\rm ell}^{3/2}m_{\rm p} =0.62(6)$\,M$_{\odot}$. 

For low-mass, slightly evolved stars with $4500\leq T_{\rm eff}\leq 6000$\,K and $3.5\leq \log g \leq 4.5$\,dex, spherically symmetric model atmospheres convolved with the Kepler bandpass (which is similar to that of our CCD) yield gravity and limb-darkening coefficients that imply  $f_{\rm ell}\simeq 1.45-1.65$ \citep[e.g.][ and references therein]{nl13}. This  would suggest a low-mass neutron star with $m_{\rm p} \lesssim 1.4$\,M$_{\odot}$. If $f_{\rm RLO} < 1$, then the pulsar mass should be even lower. 

Obviously, there are several caveats related to this analysis. For instance, the tidal deformation is treated perturbatively.  Given the $\sim 10\%$ amplitude of the signal and the almost perfectly symmetrical maxima, this should be a adequate approximation (i.e. higher-order terms should be small). In addition, the estimate for $f_{\rm ell}$ may differ substantially for non-symmetric Roche-lobe filling stars. From Eq.\,4 one sees that a massive (e.g. $1.8$\,M$_{\odot}$) would require an effective $f_{\rm ell}\geq 1.73$.

A more realistic estimate for the system's geometry can be obtained through a detailed  numerical model. We modeled the quiescent lightcurve with the \textsc{nightfall}\footnote{http://www.hs.uni-hamburg.de/DE/Ins/Per/Wichmann/Nightfall.html} lightcurve-synthesis code \citep{nightfall}, using $f_{\rm fill}$, $i$ and $T_{\rm eff}$ as free parameters, while keeping the mass-ratio fixed to our spectroscopic estimate. The flux was calculated using tabulated NextGen atmospheres with $\log g = 4.0$\,dex, integrated over the Bessel $R$ bandpass. 
This analysis yields overall consistent results, with $i=41(3)\deg$, $f_{\rm fill}>0.96$, $T_{\rm eff}=5600(600)$\,K, and $\chi^{2}_{\rm red,min}\simeq0.45$, which agree well with our semi-analytic estimate, but imply an  even lower pulsar mass at face value. The best-fit synthetic lightcurve is shown in Figure\,\ref{fig:2}b. 
In conclusion, while  both the analytic and numerical constraints for the component masses may be biased, the inferred pulsar mass is rather typical and consistent with the MSP masses observed in nature  \citep[e.g.][]{ato+16}.

\subsection{Irradiation and intra-binary shocks}
Interestingly, in {X}-rays, PSR\,J1723$-$2837  shows evidence for non-thermal emission modulated at the orbital period \cite{bec+14}. This hints to the presence of an intra-binary shock front which, if persistent, could lead to indirect heating of the companion \citep{rs16}. In \S\ref{sec:ellipsoidal} we discuss evidence for that based on a small signal with a maximum misaligned with inferior conjunction during  quiescent periods. 
Even though our estimates are likely contaminated by residual surface activity, taken at face value they still imply a relatively small irradiation efficiency ($\simeq 4-10$\% adopting a surface temperature of $\sim 5500$\,K and a spin-down luminosity of $4.6\times10^{34}$\,erg\,s$^{-1}$ from \cite{cls+13}). 
An additional significant source of uncertainty here is the kinematic contribution to $\dot{P}_{\rm s}$ for the pulsar, which thus far remains unconstrained \citep{ cls+13}.  

In summary, $f_{\rm irr}$ must be small, but  our data do not suffice provide  useful constraints. 
We do however believe that misinterpretation of time-series optical photometry may lead to overestimates for the irradiation efficiency for other systems. 

\subsection{Relation to other eclipsing MSPs}\label{sec:other}
While surface activity has not been confidently detected in other binary MSPs, in retrospect, a few systems share striking similarities with PSR\,J1723$-$2737 suggesting that this may be a common property of eclipsing MSPs. 
For instance \cite{bkb+16} report a similar secular change in average flux for PSR\,J2129$-$0429, and a sine term at the fundamental, which they discuss in the context of relativistic beaming;  \cite{drc+16} also observe a similar signal in PSR\,J1048+2339, while \cite{lht14} note that a combination of ellipsoidal variations, heating and starspot activity is required to explain the lightcurve of PSR\,J1628$-$2305. Finally, \cite{rfc15} find spectroscopic evidence for photospheric eruptions in the black-widow system PSR\,1311$-$3430.  

In addition,the long-term timing properties of several systems, including PSRs\,J1023+0038 \citep{akh+13}, J2051$-$0827 \citep{svf+16}, J2239$-$0533 \citep{pc15} and B1957+20 \citep{as94} provide further indirect evidence for magnetized secondaries \citep{app92}. 
More generally, a magnetic field would also lead to an enhanced stellar wind, thereby providing a possible mechanism to explain the radio eclipses, even in cases where the secondary is under-filling its Roche lobe and the pulsar irradiation is negligible. 

The presence of strong magnetic fields in all redback companions would not be surprising,  as rotation is believed to enhance the stellar dynamo \citep{spots}. Both young stars in open clusters and late-type rapid rotators in binaries show evidence for persistent large spots even at high latitudes and close to their polar caps \citep{spots,kor12}. Stellar rotation also influences the duration of magnetic cycles, with faster rotation often leading to shorter activity lengths \citep{sb99,wdm+11}. If those cycles  regulate the stellar wind, they could therefore lead to variable mass outflows with timescales similar to those of the MSP state transitions.

\acknowledgements
We thank the anonymous referees for their helpful and constructive reports. 
JA is a Dunlap Fellow at the Dunlap Institute for
Astronomy and Astrophysics at the University of Toronto.  The Dunlap
Institute is funded by an endowment established by the David Dunlap
family and the University of Toronto.  We have made extensive use of NASA's Astrophysics Data System.


\end{document}